\documentclass[aps,pra,showkeys,twocolumn,superscriptaddress]{revtex4-2}
\usepackage{array}
\usepackage{booktabs}
\usepackage{tabu}
\usepackage{dcolumn}
\usepackage{amsmath}
\usepackage{amsfonts}
\usepackage{float}
\usepackage{amssymb}
\usepackage{graphicx,color}
\usepackage[colorlinks={true}]{hyperref}
\hypersetup{colorlinks=true,linkcolor=blue,citecolor=blue,urlcolor=blue}
\usepackage{graphicx}
\usepackage{subfigure}
\usepackage{graphicx}
\usepackage{dcolumn}
\usepackage{bm}
\usepackage{pstricks}
\usepackage{braket}
\usepackage{orcidlink}

\def\be{\begin{equation}}
  \def\ee{\end{equation}}
\def\bea{\begin{eqnarray}}
\def\eea{\end{eqnarray}}
\def\f{\frac}
\def\n{\nonumber}
\def\l{\label}
\def\p{\phi}
\def\o{\over}
\def\R{\rho}
\def\pa{\partial}
\def\om{\omega}
\def\na{\nabla}
\def\P{\Phi}
\begin{document}

\title{Cavity-Mediated Charging of a Graphene Excitonic Quantum
Battery} 

\author{Maryam Hadipour \orcidlink{0000-0002-6573-9960}}
\email{maryam.hadipour1362@gmail.com}
\affiliation{Faculty of Physics, Urmia University of Technology, Urmia, Iran}

\author{Soroush Haseli \orcidlink{0000-0003-1031-4815}}\email{soroush.haseli@gmail.com}
\affiliation{Faculty of Physics, Urmia University of Technology, Urmia, Iran}

\date{\today}
\def\be{\begin{equation}}
  \def\ee{\end{equation}}
\def\bea{\begin{eqnarray}}
\def\eea{\end{eqnarray}}
\def\f{\frac}
\def\n{\nonumber}
\def\l{\label}
\def\p{\phi}
\def\o{\over}
\def\R{\rho}
\def\pa{\partial}
\def\om{\omega}
\def\na{\nabla}
\def\P{$\Phi$}

\begin{abstract}
We study the charging and work-extraction properties of a graphene-based excitonic quantum battery embedded in a driven-dissipative optical microcavity. The system consists of a pair of intervalley excitons in strained graphene, where one exciton acts as the charger and the other as the quantum battery, both coupled to a common cavity mode through a Tavis-Cummings interaction. By solving the open-system dynamics, we analyze the ergotropy as a measure of extractable work and investigate how coherent and incoherent pumping, cavity loss, and the microcavity parameter influence the charging process. Our results show that the battery exhibits a transient ergotropy peak followed by relaxation to a steady state, with the maximum extractable work strongly controlled by the light-matter coupling strength. The study reveals an optimal regime for efficient charging and demonstrates the role of cavity engineering in enhancing work storage in excitonic quantum batteries. 
\end{abstract}

\maketitle
\paragraph*{Introduction}.
Quantum batteries have emerged as a promising research topic in quantum thermodynamics and quantum technologies because they may exploit uniquely quantum features to improve energy storage and transfer. Early foundational studies showed that collective operations can significantly enhance charging power compared with local strategies, thereby establishing the basic framework of the field \cite{campaioli2017}. Soon after, attention expanded from charging power to extractable work, showing that entanglement and coherence can play an important role in battery performance \cite{alicki2013,Shi2022}. A major line of research has focused on mechanisms for enhancing charging speed and efficiency. Harmonic driving, noise-assisted charging, and the use of correlated chargers have all been shown to improve performance in suitable regimes \cite{zhang2019,ghosh2021,arjmandi2022}. These results indicate that the advantage of quantum batteries depends not only on the battery itself, but also on the charging protocol and the structure of the charger. Another important direction concerns open quantum batteries, where the environment is not treated only as a source of decoherence, but also as a possible resource. Environment-mediated charging, non-Markovian memory effects, and collisional models have demonstrated that environmental interactions can substantially modify charging and self-discharging dynamics \cite{tabesh2020,kamin2020,morrone2023}. Related studies have introduced bounds on charging power and shown that environment engineering, dissipative ancillas, nonequilibrium steady-state currents, and reservoir control can improve performance and stability \cite{zakavati2021,xu2021,kamin2024,kamin2023,cavaliere2025}. The role of extractable work is commonly quantified through ergotropy, which measures the maximum useful work obtainable from a quantum state. Recent studies of nonequilibrium steady states in many-body systems such as XXZ spin chains have shown that interactions and steady-state structure strongly affect battery performance \cite{mojaveri2024}. This reflects a broader shift from idealized few-body models toward more realistic many-body settings. At the same time, several physical platforms for realizing quantum batteries have been investigated, including solid-state systems, superconducting circuits, optomechanical setups, and existing quantum-computing hardware such as IBM Quantum devices \cite{ferraro2018,hu2022,shokri2025,gemme2022}. These efforts show that the field is moving gradually from theoretical proposals toward experimental implementation. Recent work has also emphasized stability and robustness. Remote charging, suppression of degradation, mitigation of self-discharging, and fast adiabatic or counter-diabatic protocols have all been proposed to balance charging power with long-term stability and control \cite{song2024,song2025,fasihi2025,moraes2024}. In parallel, nonstandard charging resources such as quantum measurements, parity-deformed fields, and catalytic auxiliary systems have broadened the range of available charging strategies \cite{du2025,mojaveri2024parity,rodriguez2023}. More recently, quantum batteries have been connected with many-body physics, phase transitions, and topology. Examples include three-level Dicke batteries, charging across quantum critical regions, and topological quantum batteries \cite{yang2024,grazi2024,lu2025}. In addition, optimal-control methods and artificial-intelligence-based approaches have been developed to design more efficient charging protocols \cite{mazzoncini2023,rodriguez2023ai}. These developments have also motivated deeper discussions of what should count as a genuine quantum advantage in battery charging \cite{rinaldi2025,andolina2025}. Overall, the literature shows that quantum-battery research has evolved from simple conceptual models to a broad and active field involving open-system dynamics, many-body effects, control theory, and experimental platforms. Review articles now emphasize both the opportunities and the remaining challenges, including scalability, stability, extractable work, self-discharging, and the identification of genuine quantum advantage \cite{quach2023,campaioli2024,ferraro2026}.

Excitons are electron-hole bound states and can be viewed as hydrogen-like quasiparticles in condensed-matter systems \cite{Kasha1965}. In a gapped graphene monolayer, these excitonic states provide a convenient platform for constructing effective two-level systems, especially when the material is placed inside an optical microcavity and exposed to a strain-generated pseudomagnetic field \cite{Guinea2010,Zhu2015}. Such a field may arise from mechanical deformation of graphene and, unlike an ordinary magnetic field, it appears with opposite orientation in the two inequivalent valleys, so the full system retains time-reversal symmetry \cite{Guinea2010,Zhu2015}. Related pseudomagnetic effects induced by optical means have also been reported in other atomically thin materials \cite{Kim2014}.

The strain-induced pseudomagnetic field strongly reshapes the excitonic spectrum. Instead of a continuous dispersion, the exciton energies become quantized into discrete Landau-like levels \cite{Berman2022}. When an optical cavity is tuned close to resonance with the transition between the lowest two excitonic levels, each exciton can be treated as an effective qubit. This makes the system suitable for describing energy exchange processes in a quantum-battery setting. In the present formulation, we consider two excitons: one plays the role of the charger, while the other serves as the quantum battery. Their interaction with a common cavity mode is described within the Tavis-Cummings framework \cite{Tavis1968}.

Because the cavity is not perfectly isolated, the problem must be treated as an open quantum system. Dissipation in the cavity and relaxation of the excited excitonic states influence the charging process and therefore affect the amount of energy that can be deposited into the battery exciton. A coherent external drive is included in the cavity in order to supply energy to the charger subsystem and to mediate energy transfer toward the battery exciton. In this way, the microcavity acts as the interface through which excitation energy is injected, redistributed, and partially lost to the environment.

This model allows one to investigate several central quantities in quantum-battery theory, such as the charging rate, stored energy, and the impact of dissipation on the charging protocol. The role of the pseudomagnetic field is particularly important because it determines the excitonic level structure and therefore controls the resonance condition with the cavity mode \cite{Berman2022}. At the same time, cavity parameters such as linewidth, mode frequency, dielectric environment, and volume influence the strength and efficiency of the energy-transfer mechanism. By tuning these quantities, one can explore how the charging performance of the excitonic battery changes under different physical conditions.

The use of a cavity-driven excitonic platform is consistent with broader developments in quantum-battery research, where effective few-level systems and collective light-matter interactions are frequently employed to model charging and storage protocols \cite{campaioli2017,alicki2013,Shi2022}. In this sense, the graphene-based setup provides a solid-state realization in which strain engineering offers an additional degree of control not usually available in more conventional battery architectures. The pseudomagnetic field does not merely perturb the system, but rather defines the structure of the excitonic states that participate in the charging dynamics.

Although the same driven-dissipative formalism has been used in earlier studies of excitonic correlations in microcavities \cite{Martins2022,Albert2013}, here it is reinterpreted from the perspective of energy storage. The physical picture is that one exciton is actively supplied with energy through the cavity field and subsequently transfers a portion of that energy to the second exciton, which functions as the storage element. This viewpoint makes the model directly relevant to the theory of quantum batteries and to the study of controlled energy flow in hybrid quantum devices.

Overall, a pair of excitons in strained graphene embedded in a leaky optical cavity forms a useful minimal model for a quantum battery. The combination of strain-induced pseudomagnetic quantization, cavity-assisted coupling, and external coherent pumping provides a flexible setting for examining how stored energy and charging efficiency depend on both material and photonic parameters. Such a framework may also be extended to larger excitonic assemblies, where one may study collective charging effects and possible enhancements of battery performance in many-body regimes.

\begin{figure}[h]
\includegraphics[width=0.5 \textwidth]{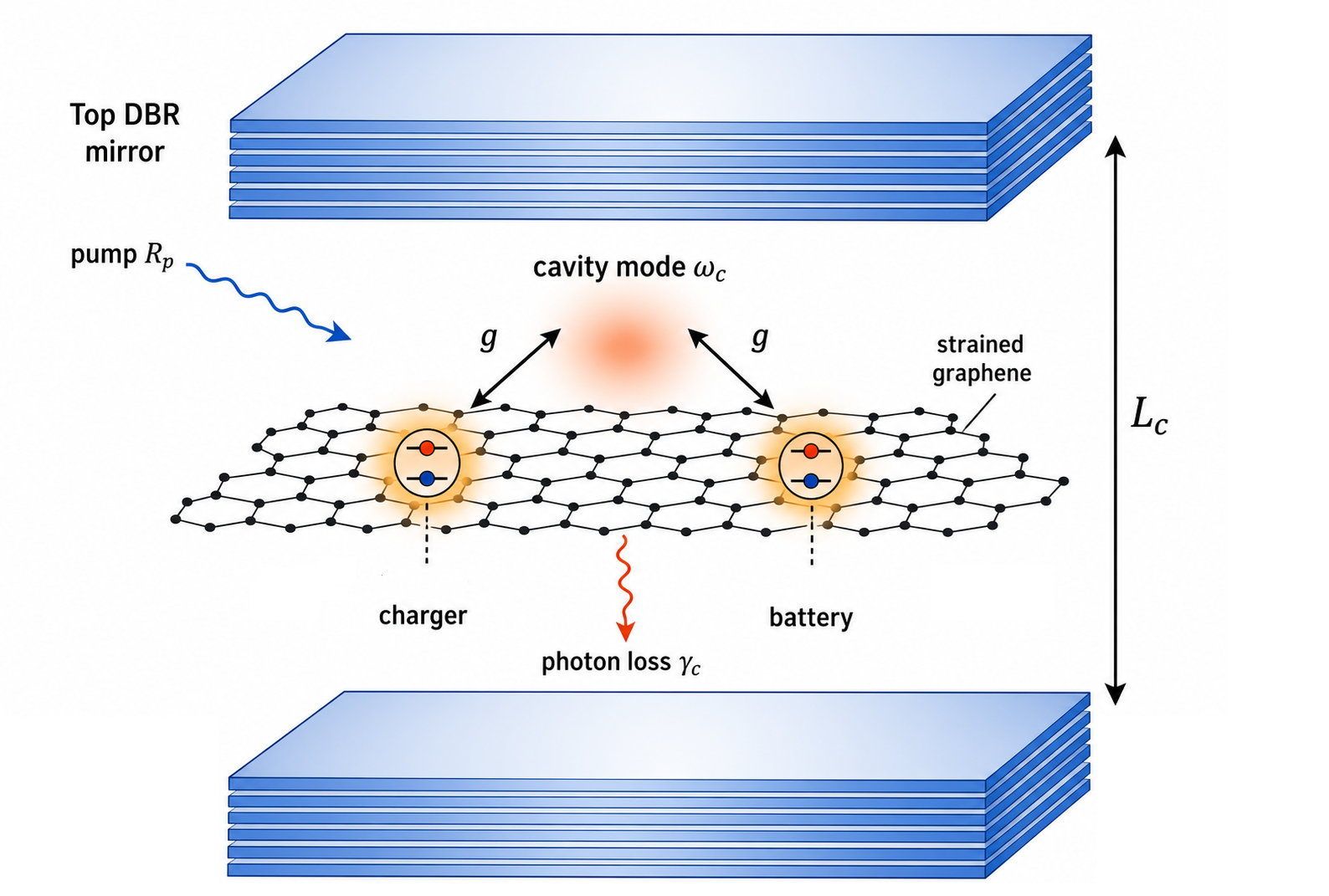}
\caption{ A strained graphene layer in a DBR microcavity of length $L_c$. Two excitonic centers, acting as charger and battery, couple to a central cavity mode $\omega_c$ with strength $g$. The system is subject to external pumping $R_p$ and cavity dissipation $\gamma_c$.}
\label{Fig1}
\end{figure}

\paragraph*{Graphene Excitonic Quantum-Battery Model}.
We consider a pair of intervalley excitons in a strained graphene monolayer placed inside a leaky optical microcavity formed by two distributed Bragg reflectors (DBRs).
The graphene sheet is located between the two mirrors, which are separated by the cavity length $L_C$. The strain applied to the graphene monolayer induces a perpendicular pseudomagnetic field $B$, which leads to the quantization of the electronic spectrum and the emergence of discrete excitonic transitions \cite{Guinea2010}. A schematic representation of this configuration is illustrated in Fig.~\ref{Fig1}. Each relevant excitonic transition is modeled as an effective two-level system, with a ground state $|0\rangle_j$ and an excited state $|1\rangle_j$, where $j=1,2$ labels the two excitons. In the low-excitation-density regime, direct exciton-exciton interactions are neglected, and the excitons interact only indirectly through their coherent coupling to a shared quantized cavity mode. Within the framework of the quantum-battery model adopted here, one exciton is designated as the charger, while another exciton  serves as the quantum battery ($QB$). These labels define the initial preparation and operational roles of the two excitons, but they do not break the symmetry of their interaction with the cavity mode.  The transition energy of a direct exciton is defined by the energy difference between its excited and ground states, expressed as
$
\Delta = \epsilon_{0,0,1}^0 - \epsilon_{0,0,0}^0 = \frac{\pi^2 k e^2}{2 \epsilon_d l},
$
,
where $k$ denotes the Coulomb constant, $e$ is the elementary charge, $\epsilon_d$ represents the dielectric constant of the intracavity medium and $l$ is the effective pseudomagnetic length \cite{Berman2022}. It is defined by
$
l = \hbar B,
$
where $B$ denotes the pseudomagnetic-field parameter in the convention used for the strained-graphene spectrum. Since $l \propto B^{-1/2}$, the excitonic transition energy scales with the physical pseudomagnetic field according to
$
\Delta \propto B^{1/4}
$.
The optical cavity is assumed to remain resonant with the excitonic transition throughout the analysis. The resonance condition is therefore
$
\hbar \omega_c = \Delta
$
, where $\omega_c$ is the angular frequency of the fundamental cavity mode. Consequently, when the pseudomagnetic field is varied, the cavity frequency is adjusted so that the cavity mode remains resonant with the excitonic transition.
Within the rotating-wave approximation, the coherent dynamics of the two excitons and the shared cavity mode are governed by the two-emitter Tavis–Cummings Hamiltonian \cite{Tavis1968,Martins2025d}
\begin{equation}\label{H-Tvis}
\hat{H}_0 = \hbar \omega_c \hat{a}^\dagger \hat{a} + \sum_{j=1}^{2} \left[ \frac{\Delta}{2} \hat{\sigma}_z^{(j)} + g_j \left( \hat{a} \hat{\sigma}_+^{(j)} + \hat{a}^\dagger \hat{\sigma}_-^{(j)} \right) \right],
\end{equation}
where $\hat{a}$ and $\hat{a}^\dagger$ are the annihilation and creation operators of the cavity mode, respectively, while $\hat{\sigma}_z^{(j)}$, $\hat{\sigma}_+^{(j)}$, and $\hat{\sigma}_-^{(j)}$ represent the Pauli inversion, raising, and lowering operators for the $j$-th exciton.  The ladder operators are defined as
$
\hat{\sigma}_+^{(j)} = |1\rangle_j \langle 0|, \quad \hat{\sigma}_-^{(j)} = |0\rangle_j \langle 1|
$
.
The interaction term $\hat{a} \hat{\sigma}_+^{(j)}$ represents the absorption of a cavity photon that excites the $j$-th exciton, while $\hat{a}^\dagger \hat{\sigma}_-^{(j)}$ describes the inverse process, in which the exciton decays by emitting a photon into the cavity mode. Both processes preserve the total number of excitations. We further assume that the two excitons are spatially equivalent with respect to the cavity field and possess identical transition dipole moments. Consequently, we assume equal coupling strengths, $g_1 = g_2 = g$. It should be noted that there is no direct interaction between the quantum battery and charger. The cavity–exciton coupling strength is given by \cite{Martins2022}
\begin{equation}
g = e v_F \sqrt{\frac{\pi \hbar}{2 \epsilon_0 \epsilon_d \Delta_{ex} W}},
\end{equation}
where $v_F \approx 10^6$ m/s is the Fermi velocity for electrons in graphene, $\epsilon_0$ is the vacuum permittivity, and $W$ represents the effective mode volume of the optical cavity.  The microcavity parameter is  introduced as $\alpha = \epsilon_d W$, which aggregates both the dielectric and geometric properties of the cavity. So, an increase in the microcavity parameter suppresses the light–matter coupling strength, whereas an increase in the pseudomagnetic field leads to a more gradual decline of this coupling. 
For a fixed pseudomagnetic field, notably for the calculations performed at $B = 50 T$ the dependence of the coupling strength on the cavity parameter can be expressed relative to a reference value $\alpha_0$ as
$
g(\alpha) = g_0 \sqrt{\frac{\alpha_0}{\alpha}},
$
,
where $g_0 = g(\alpha_0)$ denotes the coupling strength at this reference point. For instance, in studies utilizing an InGaAs microcavity with a dielectric constant $\epsilon_d \approx 13$ and mode volume $W = 1.69 \times 10^3 \, \mu\text{m}^3$,   the corresponding reference parameter is defined  as $\alpha_0 \equiv 13 \times 1.69 \, \mu\text{m}^3$ \cite{Zhang2016y}. So, tuning the dimensionless ratio $\alpha/\alpha_0$ modulates the coherent exciton–photon interaction without affecting either the exciton transition energy or the resonant cavity frequency. Because the DBR mirrors are not perfectly reflecting and the excitons may 
interact with environmental degrees of freedom, the coupled cavity-exciton 
system is treated as an open quantum system and its density operator $\hat{\rho}(t)$ evolves according to the following master equation as \cite{nilsen}
\begin{equation}
\frac{d\hat{\rho}}{dt}
=
-\frac{i}{\hbar}
\left[\hat{H}_0,\hat{\rho}\right]
+
\gamma_{c}\mathcal{D}[\hat{a}]\,\hat{\rho}
+
\gamma_{q}\sum_{j=1}^{2}
\mathcal{D}\!\left[\hat{\sigma}_{-}^{(j)}\right]\hat{\rho},
\label{eq:master_equation}
\end{equation}
where $\gamma_{c}$ denotes the cavity-photon decay rate, $\gamma_{q}$ is the 
exciton relaxation rate, and $\hat{\sigma}_{-}^{(j)}$ is the lowering operator 
of the $j$th exciton. The Lindblad dissipator associated with an arbitrary 
collapse operator $\hat{O}$ is defined as
\begin{equation}
\mathcal{D}[\hat{O}]\,\hat{\rho}
=
\hat{O}\hat{\rho}\hat{O}^{\dagger}
-\frac{1}{2}
\left\{
\hat{O}^{\dagger}\hat{O},\hat{\rho}
\right\},
\label{eq:lindblad_dissipator}
\end{equation}
where
$\{\hat{A},\hat{B}\}
=\hat{A}\hat{B}+\hat{B}\hat{A}$
denotes the anticommutator.

To investigate the performance of the graphene excitonic battery under external driving, we consider two distinct pumping mechanisms for the cavity mode: incoherent and coherent. These mechanisms differ fundamentally in the way they populate the cavity and, consequently, in how they influence the battery's work-extraction capacity.

\paragraph*{Incoherent pumping}.
Incoherent pumping is incorporated into the Lindblad master equation via the addition 
of a pump dissipator, $\gamma_{p}\mathcal{D}[\hat{a}^{\dagger}]\hat{\rho}$, where 
$\gamma_{p}$ denotes the incoherent photon injection rate. This term represents a 
phase-insensitive excitation source that increases the cavity photon population 
while simultaneously inducing statistical mixing in the system. In the context of 
the quantum battery, this mechanism allows for the replenishment of energy lost 
to cavity leakage, potentially sustaining the battery ergotropy, albeit at the 
cost of increasing the state mixedness.  Here, we investigate how an incoherent cavity-photon source influences the work extraction process of the quantum battery within the Tavis-Cummings framework. To this end, we numerically solve the master equation for the density matrix $\rho$,
\begin{equation}
\frac{d\hat{\rho}}{dt} = -\frac{i}{\hbar} \left[ \hat{H}_0, \hat{\rho} \right] + \gamma_c \mathcal{D}[\hat{a}]\hat{\rho} + \gamma_p \mathcal{D}[\hat{a}^\dagger]\hat{\rho} + \gamma_q \sum_{j=1}^{2} \mathcal{D}[\hat{\sigma}_-^{(j)}]\hat{\rho},
\label{eq:master_equation_full}
\end{equation}
where $\hat{H}_0$ is the Tavis-Cummings Hamiltonian introduced in Eq.~(\ref{H-Tvis}). In Eq.~\eqref{eq:master_equation_full}, the term $\gamma_p \mathcal{D}[\hat{a}^\dagger]\hat{\rho}$ describes the phase-insensitive incoherent injection of photons into the cavity at a rate of $\gamma_p$, while $\gamma_c$ and $\gamma_q$ represent the cavity decay rate and the exciton spontaneous emission rate, respectively. The charger-to-battery transfer protocol is initialized with the cavity in its vacuum state, the charger exciton in its excited state, and the battery exciton in its ground state. Accordingly, the total system contains exactly one excitation at the initial time. The corresponding initial state is
\begin{equation}
|\psi(0)\rangle = |0\rangle_c \otimes |1\rangle_{\mathrm{charger}} \otimes |0\rangle_{\mathrm{battery}}.
\end{equation}
To illustrate the effect of incoherent pumping on the work extraction capability of the battery, we first consider the two-excitonic-qubit system in the absence of qubit decay, i.e., $\gamma_q = 0$. The ergotropy, denoted by $\mathcal{W}(\tau)$, quantifies the maximum amount of work that can be extracted from a quantum battery (QB) under cyclic unitary operations after a charging period $\tau$ \cite{Allahverdyan2004}. Mathematically, it is defined as
\begin{equation}
    \mathcal{W}(\tau) = \text{Tr}\left[\rho_B(\tau) H_B\right] - \text{Tr}\left[\sigma_{\rho_B} H_B\right],
    \label{eq:ergotropy}
\end{equation}
where $H_B$ is the Hamiltonian of the battery, $\rho_B(\tau)$ is the state of the battery at the charging time $\tau$, and $\sigma_{\rho_B}$ represents the passive state associated with $\rho_B(\tau)$. The passive state $\sigma_{\rho_B}$ is a diagonal state (unitarily equivalent to $\rho_B$) with eigenvalues sorted in descending order corresponding to the ascending order of the energy levels of $H_B$, from which no further work can be extracted under cyclic unitary transformations.

To characterize the performance of the charging process, we introduce two key quantities. First, the maximum ergotropy $\mathcal{W}_{\max}$ represents the peak extractable work optimized over the entire charging duration:
\begin{equation}
    \mathcal{W}_{\max} = \max_{\tau} \left[ \mathcal{W}(\tau) \right].
    \label{eq:max_ergotropy}
\end{equation}
Second, we define the steady-state ergotropy, $\mathcal{W}_{\text{ss}}$, as the ergotropy attained in the long-time limit ($\tau \to \infty$) where the system reaches a stationary regime characterized by $\dot{\rho} = 0$.
\begin{figure}[h]
\includegraphics[width=0.25 \textwidth]{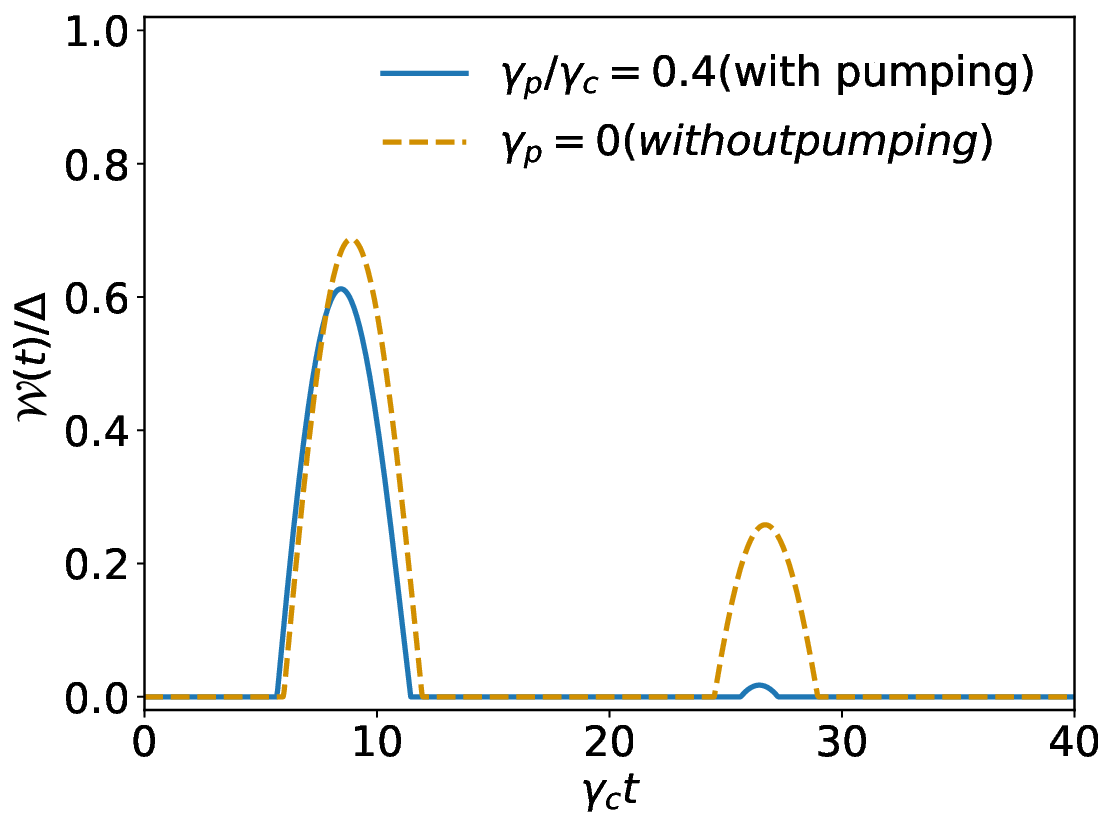}
\caption{$\mathcal{W}(t)/\Delta$ as function of $\gamma_c t$. The solid blue curve shows the dynamics under incoherent pumping $\gamma_p / \gamma_c =0.4$, while the dashed yellow curve corresponds to the case without pumping $\gamma_p=0$ for both the coupling strength is $g=0.25$.}
\label{Fig2}
\end{figure}
In Fig. (\ref{Fig2}), the normalized ergotropy $\mathcal{W}(t)/\Delta_{ex}$ 
is plotted as a function of time for the cases with and without pumping. In both cases, the ergotropy exhibits an initial transient growth, signaling the charging process of the excitonic battery. The pumped case shows a smaller maximum and stronger damping, since incoherent pumping increases the mixedness of the state and reducing its non-passive character. As a result, a smaller fraction of the stored energy can be extracted as useful work. By contrast, in the absence of pumping, the dynamics preserve coherence more effectively, leading to a larger ergotropy peak and a more pronounced revival at later times.

In contrast to incoherent pumping, coherent pumping is modeled within the 
semiclassical approximation, wherein the external driving field is treated 
classically while the cavity mode remains fully quantized \cite{Gerry}. In this approach, 
the effect of the drive is introduced directly at the Hamiltonian level, such 
that the bare Hamiltonian $\hat{H}_0$ is replaced by the total time-dependent 
Hamiltonian
$
\hat{H}=\hat{H}_0+\hat{H}_{\mathrm{drive}}(t)
$. The coherent driving term is written as
\begin{equation}
\hat{H}_{\mathrm{drive}}(t)
=
\hbar R_{P}
\left(
\hat{a}\,e^{i\omega_{L} t}
+
\hat{a}^{\dagger}e^{-i\omega_{L} t}
\right),
\end{equation}
where $R_{P}$ denotes the coherent driving amplitude and $\omega_{L}$ is the
frequency of the external laser field. The system dynamics is governed by the following master equation
\begin{equation}
\hbar\dot{\rho}
=
\frac{1}{i}\left[\hat{H},\rho\right]
+
\gamma_c\mathcal{L}(\hat{a})\rho
+
\gamma_q\sum_{j=1}^{N}
\mathcal{L}\left(\sigma_-^{j}\right)\rho.
\end{equation}
 Unless explicitly stated otherwise, all simulations are carried out under resonant conditions, so that the laser frequency equals the cavity-mode frequency, $\omega_L=\omega_c$, and the cavity mode remains resonant with the excitonic transition, $\hbar\omega_c=\Delta$.
\begin{figure}[h]
\includegraphics[width=0.45 \textwidth]{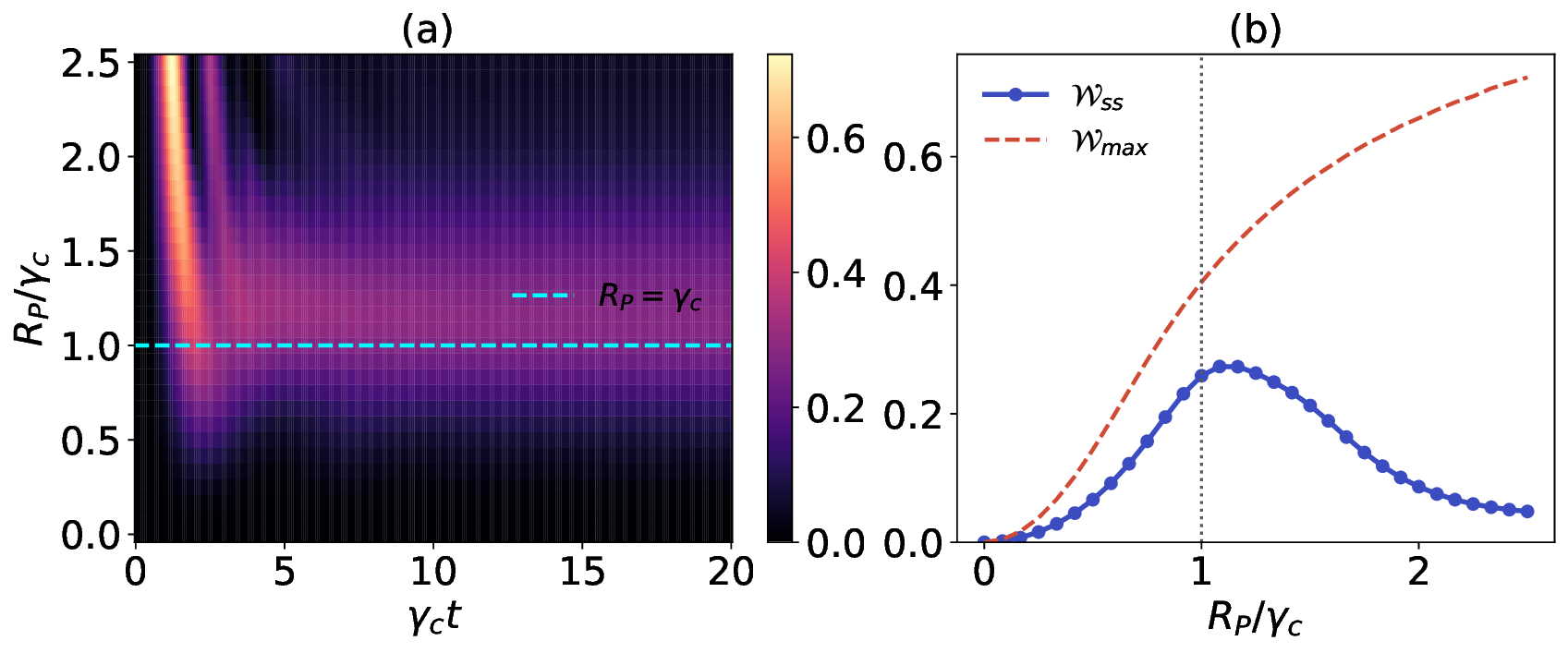}
\caption{(a) Density plot of the normalized ergotropy as a function of the scaled time $\gamma_c t$ and the normalized pumping rate $R_P/\gamma_c$. The dashed horizontal line marks the threshold condition $R_P=\gamma_c$. (b) Steady-state ergotropy, $\mathcal{W}_{\mathrm{ss}}$, and maximum ergotropy, $\mathcal{W}_{\mathrm{max}}$, plotted as functions of the normalized pumping rate $R_P/\gamma_c$. The dotted vertical line indicates $R_P/\gamma_c=1$.}\label{fig:fig3}
\end{figure}
In Fig.~\ref{fig:fig3}, we present the dynamics of the  ergotropy as a function of the scaled time $\gamma_c t$ and the normalized pumping rate $R_P/\gamma_c$. Fig.~\ref{fig:fig3}(a) shows a density plot in which the color intensity represents the magnitude of the normalized ergotropy, while Fig.~\ref{fig:fig3}(b) summarizes the corresponding steady-state ergotropy, $\mathcal{W}_{\mathrm{ss}}$, and the maximum transient ergotropy, $\mathcal{W}_{\max}$, as functions of the pumping strength.

From Fig.~\ref{fig:fig3}(a), it is clear that for weak pumping, $R_P/\gamma_c \ll 1$, the ergotropy remains very small over the entire time evolution, indicating that quantum battery stays close to a passive state and only a negligible amount of useful work can be extracted. As the pumping rate increases, a pronounced transient buildup of ergotropy emerges at short times, revealing that the charging process becomes more efficient when the coherent pumping starts to compete with the cavity dissipation. The bright regions appearing in the intermediate pumping regime indicate the existence of favorable operating points where the injected energy is converted into extractable work before dissipative effects dominate. In particular, near the threshold $R_P/\gamma_c \approx 1$, marked by the dashed horizontal line, the ergotropy becomes both stronger and more extended in time, suggesting an optimal balance between energy injection and cavity loss.

For pumping rates beyond this threshold, the transient ergotropy can still reach relatively large values at early times, but its long-time persistence is clearly reduced. This behavior indicates that although stronger pumping injects more energy into the system, it does not necessarily improve the sustained work-storage capability of the battery. Instead, excessive pumping tends to drive the system toward a more mixed stationary state, where the increase in population is no longer accompanied by a proportional increase in useful quantum ordering. As a result, part of the supplied energy becomes locked in a form that is not fully extractable as work.

This interpretation is reinforced by Fig.~\ref{fig:fig3}(b). The red dashed curve, corresponding to $\mathcal{W}_{\max}$, increases monotonically with $R_P/\gamma_c$, showing that stronger pumping enhances the highest transient work capacity achievable during the evolution. By contrast, the blue curve, representing $\mathcal{W}_{\mathrm{ss}}$, exhibits a non-monotonic dependence on the pumping strength: it increases from nearly zero, reaches a maximum slightly above $R_P/\gamma_c=1$, and then decreases as the pumping is further increased. This non-monotonic behavior confirms that the optimal operating regime for long-time battery performance lies close to the balance point between pumping and dissipation. Below this regime, the drive is too weak to generate significant ergotropy, whereas above it, the stronger pumping promotes decoherence and mixedness, which reduce the amount of extractable work stored in the second qubit. Overall, Fig.~\ref{fig:fig3} demonstrates a clear distinction between transient and steady-state charging performance. Although increasing the pumping rate always improves the maximum ergotropy attainable at some intermediate time, the steady-state ergotropy is optimized only around $R_P/\gamma_c \approx 1$. Therefore, the figure highlights the existence of an optimal pumping window for efficient and sustainable quantum battery operation, where the competition between coherent energy injection and cavity dissipation maximizes the useful stored work while avoiding the detrimental effects of overdriving.
\begin{figure}[h]
\includegraphics[width=0.45 \textwidth]{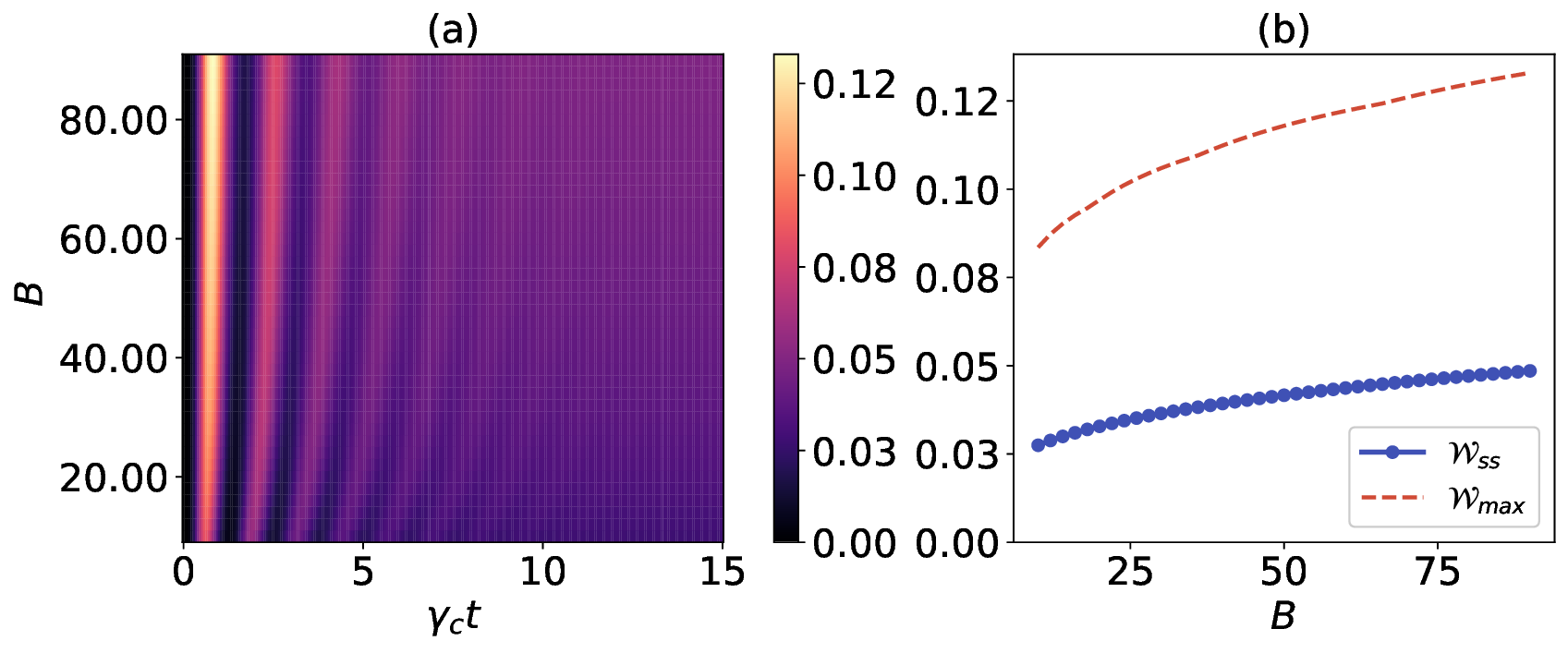}
\caption{ (a)  Density plot of $\mathcal{W}(t)/\Delta$ as function of $\gamma_c t$ and  pseudomagnetic field $B$ . (b)   $\mathcal{W}_{\rm ss}$ and $\mathcal{W}_{\rm max}$  versus pseudomagnetic field $B$. The choosen parameters are $\gamma_c=1.5$, $R_P=1.2$, $\gamma_q=0.0$ and $g_0=3.0$.}
\label{fig:ergotropy_dynamics}
\end{figure}
In Fig.(\ref{fig:ergotropy_dynamics}), we investigate the performance of the excitonic quantum battery by analyzing the dimensionless ergotropy as a function of the pseudomagnetic field (PMF) strength $B$ and the dimensionless time $\gamma_c t$. 

In Fig.(\ref{fig:ergotropy_dynamics})(a), the temporal evolution of the ergotropy illustrates the transition from the initial state to a driven-dissipative steady state. At early times, i.e., for $\gamma_c t < 5$, pronounced quantum oscillations are observed. These oscillations arise from the coherent energy exchange between the cavity mode and the two-level excitonic emitters, reflecting Rabi-like dynamics in the charging process. As the PMF strength $B$ increases, both the frequency and amplitude of these oscillations are modified due to the $B$-dependent coupling strength $g(B)$ and the excitonic gap $\Delta$. The damping of the oscillatory behavior at longer times is caused by cavity dissipation with rate $\gamma_c$, which gradually suppresses coherence and drives the system toward a non-equilibrium steady state.

In Fig.(\ref{fig:ergotropy_dynamics})(b), we compare the maximum transient ergotropy, $\mathcal{W}_{\max}$, with the steady-state ergotropy, $\mathcal{W}_{\mathrm{ss}}$. Both quantities increase monotonically with the pseudomagnetic field $B$, indicating that stronger PMF enhances the work-extractable capacity of the battery. This behavior suggests that the PMF effectively strengthens the excitonic response and improves the charging performance of the system. The difference between $\mathcal{W}_{\max}$ and $\mathcal{W}_{\mathrm{ss}}$ reflects the role of transient coherent dynamics: while the battery reaches its highest work-extraction capability during the early charging stage, a non-zero fraction of ergotropy remains available in the steady state. Therefore, increasing $B$ not only enhances the transient charging peak but also boosts the long-time stored useful energy, demonstrating the PMF as an effective control parameter for optimizing quantum battery performance.
\begin{figure}[h]
\includegraphics[width=0.45 \textwidth]{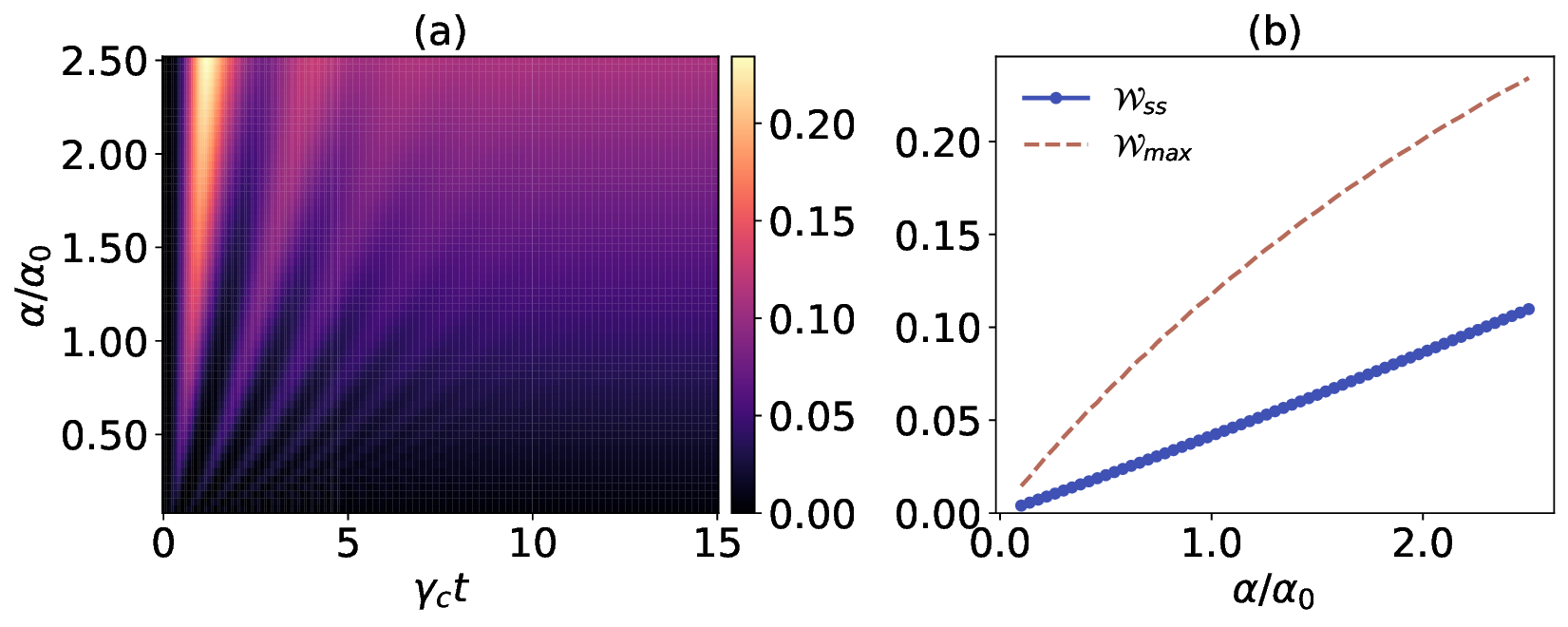}
\caption{(a) $\mathcal{W}(t)/\Delta$ versus the microcavity parameter $\alpha/\alpha_0$ and $\gamma_c t$. (b)  $\mathcal{W}_{\rm ss}$ and  $\mathcal{W}_{\rm max}$ as functions of $\alpha/\alpha_0$. The choosen parameters are $\gamma_c=1.5$, $\gamma_q=0$, $R_P=1.2$,$B=50~\mathrm{T}$ and $g_0=3.0$.}

\label{fig:ergotropy_dynamicsalpha}
\end{figure}
In Fig.~\ref{fig:ergotropy_dynamicsalpha}, the ergotropy of the excitonic quantum battery is shown as a function of the microcavity parameter $\alpha/\alpha_0$ and the scaled time $\gamma_c t$, illustrating how the charging dynamics depend on the cavity-induced coupling. Since the ergotropy quantifies the maximum extractable work from the battery state, its growth in Fig.\ref{fig:ergotropy_dynamicsalpha}~(a) reflects the build-up of useful nonequilibrium excitations under coherent pumping. The early-time rise and transient ridge indicate a charging stage where the battery is driven away from passivity, while the later saturation shows the competition between pumping, coherent light-matter coupling, and cavity dissipation. Because the coupling strength scales as $g \propto 1/\sqrt{\alpha/\alpha_0}$, changing $\alpha$ tunes the efficiency of energy transfer between the cavity field and the excitonic subsystem. For smaller $\alpha/\alpha_0$, stronger coupling generally enhances the charging rate, whereas larger $\alpha/\alpha_0$ weakens the interaction and reduces the stored work. Fig.\ref{fig:ergotropy_dynamicsalpha}~(b) summarizes this behavior by comparing the steady-state ergotropy and the maximum transient ergotropy, showing that the battery can temporarily store more extractable work than it retains at long times. This difference is a clear signature of dissipative relaxation: after the optimal charging moment, part of the non-passive energy is lost and the system settles into a mixed steady state with lower ergotropy. Overall, the figure indicates that the microcavity parameter controls an optimal charging regime where the battery reaches its best work-storage performance.

\paragraph*{Conclusion}.
In this work, we investigated the charging and work-extraction dynamics of a graphene-based excitonic quantum battery embedded in a driven-dissipative optical microcavity. The system comprises a pair of intervalley excitons in strained graphene, where one exciton functions as the charger and the other as the battery, both interacting symmetrically with a common cavity mode via a Tavis-Cummings interaction. By solving the open-system master equation, we quantified the battery's performance using ergotropy, which measures the maximum extractable work from the excitonic state.

Our results reveal that the charging dynamics are fundamentally governed by the competition between external driving, light-matter coupling, and cavity dissipation. Comparing incoherent and coherent pumping regimes, we demonstrated that while incoherent driving compensates for photon leakage, it simultaneously increases the system's mixedness, thereby reducing the ergotropy and limiting work extractability. In contrast, coherent pumping facilitates a more controlled charging process, with an optimal regime for work storage appearing near resonance. 

Furthermore, we found that the system's performance is highly tunable via material and cavity engineering. The strain-induced pseudomagnetic field and the microcavity parameter directly regulate the light-matter coupling strength, allowing for precise control over the charging rate and steady-state work storage. Our findings indicate that both the maximum transient ergotropy and the steady-state ergotropy are sensitive to these parameters, highlighting that increased coupling—achievable through specific strain configurations—significantly enhances battery performance. In conclusion, the strained-graphene excitonic platform offers a robust and controllable architecture for quantum energy storage. By carefully balancing cavity losses and pumping conditions, this architecture provides a promising pathway toward the realization of efficient, tunable, and cavity-assisted quantum batteries.



\appendix*


\begin{thebibliography}{4}
\bibitem{campaioli2017}
F. Campaioli, F. A. Pollock, F. C. Binder, L. Celeri, J. Goold, S. Vinjanampathy, and K. Modi,
Enhancing the Charging Power of Quantum Batteries,
\href{https://doi.org/10.1103/PhysRevLett.118.150601}{Phys. Rev. Lett. \textbf{118}, 150601 (2017)}.

\bibitem{alicki2013}
R. Alicki and M. Fannes,
Entanglement boost for extractable work from ensembles of quantum batteries,
\href{https://doi.org/10.1103/PhysRevE.87.042123}{Phys. Rev. E \textbf{87}, 042123 (2013)}.

\bibitem{Shi2022}
H.-L. Shi, S. Ding, Q.-K. Wan, X.-H. Wang, and W.-L. Yang,
Entanglement, Coherence, and Extractable Work in Quantum Batteries,
\href{https://doi.org/10.1103/PhysRevLett.129.130602}{Phys. Rev. Lett. \textbf{129}, 130602 (2022)}.

\bibitem{zhang2019}
Y.-Y. Zhang, T.-R. Yang, L. Fu, and X. Wang,
Powerful harmonic charging in a quantum battery,
\href{https://doi.org/10.1103/PhysRevE.99.052106}{Phys. Rev. E \textbf{99}, 052106 (2019)}.

\bibitem{ghosh2021}
S. Ghosh, T. Chanda, S. Mal, and A. Sen(De),
Fast charging of a quantum battery assisted by noise,
\href{https://doi.org/10.1103/PhysRevA.104.032207}{Phys. Rev. A \textbf{104}, 032207 (2021)}.

\bibitem{arjmandi2022}
M. B. Arjmandi, A. Shokri, E. Faizi, and H. Mohammadi,
Performance of quantum batteries with correlated and uncorrelated chargers,
\href{https://doi.org/10.1103/PhysRevA.106.062609}{Phys. Rev. A \textbf{106}, 062609 (2022)}.

\bibitem{tabesh2020}
F. T. Tabesh, F. H. Kamin, and S. Salimi,
Environment-mediated charging process of quantum batteries,
\href{https://doi.org/10.1103/PhysRevA.102.052223}{Phys. Rev. A \textbf{102}, 052223 (2020)}.

\bibitem{kamin2020}
F. H. Kamin, F. T. Tabesh, S. Salimi, F. Kheirandish, and A. C. Santos,
Non-Markovian effects on charging and self-discharging process of quantum batteries,
\href{https://doi.org/10.1088/1367-2630/ab9ee2}{New J. Phys. \textbf{22}, 083007 (2020)}.

\bibitem{morrone2023}
D. Morrone, M. A. C. Rossi, A. Smirne, and M. G. Genoni,
Charging a quantum battery in a non-Markovian environment: a collisional model approach,
\href{https://doi.org/10.1088/2058-9565/accca4}{Quantum Sci. Technol. \textbf{8}, 035007 (2023)}.

\bibitem{zakavati2021}
S. Zakavati, F. T. Tabesh, and S. Salimi,
Bounds on charging power of open quantum batteries,
\href{https://doi.org/10.1103/PhysRevE.104.054117}{Phys. Rev. E \textbf{104}, 054117 (2021)}.

\bibitem{xu2021}
K. Xu, H.-J. Zhu, G.-F. Zhang, and W.-M. Liu,
Enhancing the performance of an open quantum battery via environment engineering,
\href{https://doi.org/10.1103/PhysRevE.104.064143}{Phys. Rev. E \textbf{104}, 064143 (2021)}.

\bibitem{kamin2024}
F. H. Kamin, S. Salimi, and M. B. Arjmandi,
Steady-state charging of quantum batteries via dissipative ancillas,
\href{https://doi.org/10.1103/PhysRevA.109.022226}{Phys. Rev. A \textbf{109}, 022226 (2024)}.

\bibitem{kamin2023}
F. H. Kamin, Z. Abuali, H. Ness, and S. Salimi,
Quantum battery charging by nonequilibrium steady-state currents,
\href{https://doi.org/10.1088/1751-8121/acdb11}{J. Phys. A: Math. Theor. \textbf{56}, 275302 (2023)}.

\bibitem{cavaliere2025}
F. Cavaliere, G. Gemme, G. Benenti, D. Ferraro, and M. Sassetti,
Dynamical blockade of a reservoir for optimal performances of a quantum battery,
\href{https://doi.org/10.1038/s42005-025-01993-7}{Commun. Phys. \textbf{8}, 76 (2025)}.

\bibitem{mojaveri2024}
B. Mojaveri, R. Jafarzadeh Bahrbeig, and M. A. Fasihi,
Extracting ergotropy from nonequilibrium steady states of an XXZ spin-chain quantum battery,
\href{https://doi.org/10.1103/PhysRevA.109.042619}{Phys. Rev. A \textbf{109}, 042619 (2024)}.

\bibitem{ferraro2018}
D. Ferraro, M. Campisi, G. M. Andolina, V. Pellegrini, and M. Polini,
High-Power Collective Charging of a Solid-State Quantum Battery,
\href{https://doi.org/10.1103/PhysRevLett.120.117702}{Phys. Rev. Lett. \textbf{120}, 117702 (2018)}.

\bibitem{hu2022}
C.-K. Hu et al.,
Optimal charging of a superconducting quantum battery,
\href{https://doi.org/10.1088/2058-9565/ac8441}{Quantum Sci. Technol. \textbf{7}, 045018 (2022)}.

\bibitem{shokri2025}
A. Shokri, E. Faizi, and M. B. Arjmandi,
Entanglement-assisted charging of quantum batteries within optomechanical framework,
\href{https://doi.org/10.1103/PhysRevE.111.064117}{Phys. Rev. E \textbf{111}, 064117 (2025)}.

\bibitem{gemme2022}
G. Gemme, M. Grossi, D. Ferraro, S. Vallecorsa, and M. Sassetti,
IBM Quantum Platforms: A Quantum Battery Perspective,
\href{https://doi.org/10.3390/batteries8050043}{Batteries \textbf{8}, 43 (2022)}.

\bibitem{song2024}
W.-L. Song, H.-B. Liu, B. Zhou, W.-L. Yang, and J.-H. An,
Remote Charging and Degradation Suppression for the Quantum Battery,
\href{https://doi.org/10.1103/PhysRevLett.132.090401}{Phys. Rev. Lett. \textbf{132}, 090401 (2024)}.

\bibitem{song2025}
W.-L. Song, J.-L. Wang, B. Zhou, W.-L. Yang, and J.-H. An,
Self-Discharging Mitigated Quantum Battery,
\href{https://doi.org/10.1103/PhysRevLett.135.020405}{Phys. Rev. Lett. \textbf{135}, 020405 (2025)}.

\bibitem{fasihi2025}
M. A. Fasihi, R. Jafarzadeh Bahrbeig, B. Mojaveri, and R. Haji Mohammadzadeh,
Fast stable adiabatic charging of open quantum batteries,
\href{https://doi.org/10.1103/PhysRevE.112.024117}{Phys. Rev. E \textbf{112}, 024117 (2025)}.

\bibitem{moraes2024}
L. F. C. de Moraes et al.,
Quantum battery supercharging via counter-diabatic dynamics,
\href{https://doi.org/10.1088/2058-9565/ad71ed}{Quantum Sci. Technol. \textbf{9}, 045033 (2024)}.

\bibitem{du2025}
J. Du, Y. Guo, and B. Li,
Nonequilibrium quantum battery based on quantum measurements,
\href{https://doi.org/10.1103/PhysRevResearch.7.013151}{Phys. Rev. Research \textbf{7}, 013151 (2025)}.

\bibitem{mojaveri2024parity}
B. Mojaveri, R. Jafarzadeh Bahrbeig, and M. A. Fasihi,
Charging a quantum battery mediated by parity-deformed fields,
\href{https://doi.org/10.1103/PhysRevE.110.064107}{Phys. Rev. E \textbf{110}, 064107 (2024)}.

\bibitem{rodriguez2023}
R. R. Rodriguez et al.,
Catalysis in charging quantum batteries,
\href{https://doi.org/10.1103/PhysRevA.107.042419}{Phys. Rev. A \textbf{107}, 042419 (2023)}.

\bibitem{yang2024}
D.-L. Yang, F.-M. Yang, and F.-Q. Dou,
Three-level Dicke quantum battery,
\href{https://doi.org/10.1103/PhysRevB.109.235432}{Phys. Rev. B \textbf{109}, 235432 (2024)}.

\bibitem{grazi2024}
R. Grazi et al.,
Controlling Energy Storage Crossing Quantum Phase Transitions in an Integrable Spin Quantum Battery,
\href{https://doi.org/10.1103/PhysRevLett.133.197001}{Phys. Rev. Lett. \textbf{133}, 197001 (2024)}.

\bibitem{lu2025}
Z.-G. Lu, G. Tian, X.-Y. Lu, and C. Shang,
Topological Quantum Batteries,
\href{https://doi.org/10.1103/PhysRevLett.134.180401}{Phys. Rev. Lett. \textbf{134}, 180401 (2025)}.

\bibitem{mazzoncini2023}
F. Mazzoncini et al.,
Optimal control methods for quantum batteries,
\href{https://doi.org/10.1103/PhysRevA.107.032218}{Phys. Rev. A \textbf{107}, 032218 (2023)}.

\bibitem{rodriguez2023ai}
C. Rodriguez, D. Rosa, and J. Olle,
Artificial intelligence discovery of a charging protocol in a micromaser quantum battery,
\href{https://doi.org/10.1103/PhysRevA.108.042618}{Phys. Rev. A \textbf{108}, 042618 (2023)}.

\bibitem{rinaldi2025}
D. Rinaldi, R. Filip, D. Gerace, and G. Guarnieri,
Reliable quantum advantage in quantum battery charging,
\href{https://doi.org/10.1103/PhysRevA.112.012205}{Phys. Rev. A \textbf{112}, 012205 (2025)}.

\bibitem{andolina2025}
G. M. Andolina, V. Stanzione, V. Giovannetti, and M. Polini,
Genuine Quantum Advantage in Anharmonic Bosonic Quantum Batteries,
\href{https://doi.org/10.1103/PhysRevLett.134.240403}{Phys. Rev. Lett. \textbf{134}, 240403 (2025)}.

\bibitem{quach2023}
J. Q. Quach, G. Cerullo, and T. Virgili,
Quantum batteries: The future of energy storage?,
Future Energy \textbf{7}, 2195 (2023).

\bibitem{campaioli2024}
F. Campaioli, S. Gherardini, J. Q. Quach, M. Polini, and G. M. Andolina,
Colloquium: Quantum batteries,
\href{https://doi.org/10.1103/RevModPhys.96.031001}{Rev. Mod. Phys. \textbf{96}, 031001 (2024)}.

\bibitem{ferraro2026}
D. Ferraro, F. Cavaliere, M. G. Genoni, G. Benenti, and M. Sassetti,
Opportunities and challenges of quantum batteries,
\href{https://doi.org/10.1038/s42254-025-00906-5}{Nat. Rev. Phys. \textbf{8}, 115 (2026)}.

\bibitem{Kasha1965}
M. Kasha, H. R. Rawls, and M. Ashraf El-Bayoumi,
The exciton model in molecular spectroscopy,
Pure Appl. Chem. \textbf{11}, 371 (1965).

\bibitem{Guinea2010}
F. Guinea, M. I. Katsnelson, and A. K. Geim,
Energy gaps and a zero-field quantum Hall effect in graphene by strain engineering,
\href{https://doi.org/10.1038/nphys1420}{Nat. Phys. \textbf{6}, 30 (2010)}.

\bibitem{Zhu2015}
S. Zhu, J. A. Stroscio, and T. Li,
Programmable extreme pseudomagnetic fields in graphene by a uniaxial stretch,
\href{https://doi.org/10.1103/PhysRevLett.115.245501}{Phys. Rev. Lett. \textbf{115}, 245501 (2015)}.

\bibitem{Kim2014}
J. Kim, X. Hong, C. Jin, S. F. Shi, C. Y. S. Chang, M. H. Chiu, L. J. Li, and F. Wang,
Ultrafast generation of pseudo-magnetic field for valley excitons in WSe$_2$ monolayers,
\href{https://doi.org/10.1126/science.1258117}{Science \textbf{346}, 1205 (2014)}.
\bibitem{Berman2022}
O. L. Berman, R. Y. Kezerashvili, Y. E. Lozovik, and K. Ziegler,
Strain-induced quantum Hall phenomena of excitons in graphene,
\href{https://doi.org/10.1038/s41598-022-06917-2}{Sci. Rep. \textbf{12}, 2950 (2022)}.
\bibitem{Tavis1968}
M. Tavis and F. W. Cummings,
Exact solution for an $N$-molecule-radiation-field Hamiltonian,
\href{https://doi.org/10.1103/PhysRev.170.379}{Phys. Rev. \textbf{170}, 379 (1968)}.



\bibitem{Martins2022}
G. P. Martins, O. L. Berman, G. Gumbs, and Y. E. Lozovik,
Quantum entanglement between excitons in two-dimensional materials,
\href{https://doi.org/10.1103/PhysRevB.106.104304}{Phys. Rev. B \textbf{106}, 104304 (2022)}.

\bibitem{Albert2013}
F. Albert, K. Sivalertporn, J. Kasprzak, M. Strau\ss, C. Schneider, S. H\"ofling, M. Kamp, A. Forchel, S. Reitzenstein, E. A. Muljarov, and W. Langbein,
Microcavity controlled coupling of excitonic qubits,
\href{https://doi.org/10.1038/ncomms2764}{Nat. Commun. \textbf{4}, 1747 (2013)}.

\bibitem{Martins2025d}
G. P. Martins, O. L. Berman, G. Gumbs, and Y. E. Lozovik,
Long-lived quantum entanglement of multiple qubits: Excitons in strained graphene,
\href{https://doi.org/10.1103/PhysRevB.111.045425}{Phys. Rev. B \textbf{111}, 045425 (2025)}.
\bibitem{Zhang2016y}Q. Zhang, M. Lou, X. Li, J. L. Reno, W. Pan, J. D. Watson, M.
J. Manfra, and J. Kono, Collective non-perturbative coupling of
2D electrons with high-quality-factor terahertz cavity photons, \href{https://www.nature.com/articles/nphys3850}{Nat. Phys. 12, 1005 (2016)}.
\bibitem{nilsen}M. A. Nielsen and I. L. Chuang, Quantum Computation
and Quantum Information (Cambridge University Press, Cambridge,
2000).
\bibitem{Gerry}C. Gerry and P. Knight, Introductory Quantum Optics (Cambridge
University Press, New York, 2004).
\bibitem{Allahverdyan2004}A. E. Allahverdyan, R. Balian and Th. M. Nieuwenhuizen, Maximal work extraction from finite quantum systems,\href{https://iopscience.iop.org/article/10.1209/epl/i2004-10101-2}{EPL 67, 565 (2004)}.

\end{thebibliography}
\end{document}